\documentclass{jfm1}
\usepackage{graphicx}
\usepackage{mathrsfs}
\usepackage[citation-order]{amsrefs}

\usepackage{amsmath, amssymb, graphics}
\usepackage{latexsym, amsfonts, amssymb, txfonts, pxfonts, wasysym}
\usepackage{xcolor}

\RequirePackage{amssymb}

\ifCUPmtlplainloaded \else
  \checkfont{eurm10}
  \iffontfound
    \IfFileExists{upmath.sty}
      {\typeout{^^JFound AMS Euler Roman fonts on the system,
                   using the 'upmath' package.^^J}%
       \usepackage{upmath}}
      {\typeout{^^JFound AMS Euler Roman fonts on the system, but you
                   dont seem to have the}%
       \typeout{'upmath' package installed. JFM.cls can take advantage
                 of these fonts,^^Jif you use 'upmath' package.^^J}%
      }
  \else
  \fi
\fi

\ifCUPmtlplainloaded \else
  \checkfont{msam10}
  \iffontfound
    \IfFileExists{amssymb.sty}
      {\typeout{^^JFound AMS Symbol fonts on the system, using the
                'amssymb' package.^^J}%
       \usepackage{amssymb}%
         \let\leq=\leqslant
         \let\geq=\geqslant
      }{}
  \fi
\fi

\ifCUPmtlplainloaded \else
  \IfFileExists{amsbsy.sty}
    {\typeout{^^JFound the 'amsbsy' package on the system, using it.^^J}%
     \usepackage{amsbsy}}
    {}
\fi

\def\Real{\mbox{Re}}      
\def\Imag{\mbox{Im}}      
\def\Ai{\mbox{\rm Ai}}    

\def\sgn{\mbox{\,\rm sgn\,}}  
\def\sech{\mbox{\,sech}}  
\def\csch{\mbox{\,csch}}  
\def\hexnumber#1{\ifcase#1 0\or1\or2\or3\or4\or5\or6\or7\or8\or9\or
 A\or B\or C\or D\or E\or F\fi }
\makeatletter
\ifx\CUP@mtlplain@loaded\undefined
\else
\fi
\makeatother
\newsavebox{\thalfbox}
\sbox{\thalfbox}{$\textstyle\frac{1}{2}$}

\newsavebox{\shalfbox}
\sbox{\shalfbox}{$\scriptstyle\frac{1}{2}$}

\newsavebox{\squartbox}
\sbox{\squartbox}{$\frac{1}{4}$} 

\newsavebox{\etbox}
\sbox{\etbox}{\boldmath$\eta$}

\newsavebox{\astrutbox}
\sbox{\astrutbox}{\rule[-5pt]{0pt}{20pt}}

\mathchardef\varLambda="0103
\makeatletter
\ifx\CUP@mtlplain@loaded\undefined
\else
\fi
\makeatother
\makeatletter
\ifx\CUP@mtlplain@loaded\undefined
  \font\tenbms=cmbsy10
  \font\sevenbms=cmbsy10 at 7pt
  \font\fivebms=cmbsy10 at 5pt
  \newfam\bmsfam
  \textfont\bmsfam=\tenbms
  \scriptfont\bmsfam=\sevenbms
  \scriptscriptfont\bmsfam=\fivebms
  
  \edef\bsy@{\hexnumber\bmsfam}
  \mathchardef\bnabla="0\bsy@72
\fi
\makeatother

\font\msym=msym10
\newfam\msfam  \textfont\msfam=\msym

\newcommand{\third}{\mbox{$\!\!$\raisebox{.1ex}{
       {\footnotesize \raisebox{.6ex}{1}$\!$/$\!$\raisebox{-.6ex}{3}}}}}

\newcommand{\half}{\mbox{$\!\!$\raisebox{.1ex}{
       {\footnotesize \raisebox{.6ex}{1}$\!$/$\!$\raisebox{-.6ex}{2}}}}}

\newcommand{\beqa}{\begin{equation}}
\newcommand{\eeqa}{\end{equation}}
\newcommand{\no}{\nonumber}
\newcommand{\be}{\begin{eqnarray}}
\newcommand{\en}{\end{eqnarray}}

\newcommand{\df}[2]{\displaystyle\frac{#1}{#2}}
\newcommand{\tf}[2]{\textstyle\frac{#1}{#2}}

\newcommand{\Int}[2]{\displaystyle\int_{#1}^{#2}}

\newcommand{\PDD}[2]{\df{\partial #1}{\partial #2}}

\newcommand{\eps}{\varepsilon}

\newcommand{\Half}{\mbox{\tiny $\tf{1}{2}$}}

\title[Analytical solution of viscous KdV equation]
{Exact analytical solution of viscous Korteweg-deVries equation for water waves}

\author[S.G. Sajjadi and T.A. Smith]%
{S.\ls G.\ns S\ls A\ls J\ls J\ls A\ls D\ls I$^{1,2}$\ns\and\ns T.\ls A.\ns S\ls M\ls I\ls T\ls H$^{1}$}

\affiliation{$^{1}$\,\,\,Department of Mathematics, ERAU, Florida, USA.\\
$^{2}$\,\,\,Trinity College, University of Cambridge, UK.}

\pubyear{2014} \volume{268} \pagerange{1--30}
\date{18 August 2015}
\begin{document}
\maketitle
\begin{abstract}
\footnotesize{The evolution of a solitary wave with very weak nonlinearity which was originally investigated by Miles [4] is revisited. The solution for a one-dimensional gravity wave in  a water of uniform depth is considered. This leads to finding the solution to a Korteweg-de Vries (KdV) equation in  which the nonlinear term is small. Also considered is the asymptotic solution of the linearized KdV equation both analytically and numerically. As in Miles [4], the asymptotic solution of the KdV equation for both linear and weakly nonlinear case is found using the method of inverse-scattering theory. Additionally investigated is the analytical solution of viscous-KdV equation which reveals the formation of the Peregrine soliton that decays to the initial $\sech^2(\xi)$ soliton and eventually growing back to a narrower and higher amplitude bifurcated Peregrine-type soliton.}
\end{abstract}
\section{\sc Introduction}
The first attempt to study the effect of dissipation and dispersion was made by Chester [1] who was considering a theory for oscillations of a liquid tank near resonant frequency.  The first formal formulations of the Korteweg-deVries (KdV) equation modified by viscosity were presented by Ott and Sudan [2].  Here, the authors considered the modification which included only the dissipation of the KdV equation and electron Landau damping in plasma physics, as well as shallow water waves damped by viscosity.  The latter comprises the main thrust of the present contribution.

The most plausible formulation of the KdV equation modified by viscosity was given by Miles [3].  In his formulation, he includes the effect of both dispersion and dissipation.  Miles was motivated by his own study of the evolution of a solitary wave for very weak non-linearity [4] in order to resolve the `Ursell paradox'.  In his paper he makes a striking remark that `the viscous damping could prove more significant than non-linearity over the long time interval.'
For problems in water waves, the viscous term that appears in the KdV equations of the form [5]
\be 
u_{\tau}+c_1 uu_{\xi}+c_2 u_{\xi\xi\xi}
=c_3\int^{\infty}_{\xi}\PDD{u}{\zeta}\df{d\zeta}{\sqrt{\zeta-\xi}}\label{int1}
\en
and for an internal solitary wave, or waves in a channel of uniform but arbitrary cross section (\ref{int1}) may be modified according to [6]
\be 
{\rm LHS}[(\ref{int1})]=c_3\int^{\infty}_{-\infty}\PDD{u}{\zeta}\df{1-\sgn(\xi-\zeta)}
{\sqrt{|\xi-\zeta|}}\,d\zeta\label{int2}
\en 

Here, we shall follow Miles [3] formulation and present a closed form analytical solution to a viscous-KdV equation which manifests very intriguing results. 
However, we first give a brief account of Miles [4] theory who used the inverse scattering method to obtain the solution of the linearized KdV equation.  In his paper, Miles [4] did not present graphical results that he obtained in solving his solution of the linearized version of the KdV equation. We shall follow Miles [4] and revisit his problem and demonstrate graphically the relevant importance of the linear and nonlinear terms in the KdV equation. The graphical solutions reveal (which were omitted by Miles) some very interesting features, which we shall discuss in sections 2 and 3. 

For the purpose of demonstration we will also investigate the solution of linearized and weakly nonlinear KdV through numerical integration of respective equations.
But, the main aim of this paper is to study the viscous-KdV (VKdV) equation for surface water waves of finite depth and provide an exact analytical solution,
which motivated us by the remark made in Miles' [4] conclusion.

The results of the exact analytical solution of VKdV, taking  into account both dispersive and dissipative terms, show very interesting and surprising results. For example the formation and decay of the well known {\it Peregrine soliton}. To the best of our knowledge such a study has not been reported in the open literature.

\section{\sc Scattering formulation of KdV equation}

Miles [4] considered the solution of the KdV equation 
\be
\eta_{\tau}+\tf{1}{3}\eta_{\xi\xi\xi}+4\varepsilon\eta\eta_{\xi}=0\label{eq4}
\en
in the limit of $\eps=0$, thereby linearizing (\ref{eq4}) and reducing it to
\be
\eta_{\tau}+\tf{1}{3}\eta_{\xi\xi\xi}=0\label{eq8}
\en
subject to the initial condition 
\be
\eta(\xi,0)=\eta_0(\xi)=\eta_0(-\xi),\label{eq5}
\en
Here, 
$$
\eps=3aL^2/4d^3,
$$
where $\eps$ is the Stokes parameter, $a$ is a characteristic amplitude, and $d$ is the water depth.

The solution of equation (\ref{eq8}) may be obtained in the following form 
\be 
\eta(\xi,\tau)&=&\tau^{-\third}\int^{\infty}_{-\infty}\eta_0(\zeta)\Ai\left\{
\df{\xi-\zeta}{\tau^{\third}}\right\}\,d\zeta\quad(\eps=0)\no\\
&\sim&
\langle\eta_0\rangle\tau^{-\third}\Ai\left(\tau^{-\third}\xi\right)\hspace*{2cm}(\tau
\uparrow \infty),\label{eq9b} 
\en
with 
\be \langle\eta_0\rangle\equiv
\int^{\infty}_{-\infty}\eta_0(x)\,dx=\int^{\infty}_{-\infty}\eta(\xi,\tau)\,d\xi;\qquad \eta_0(x)=\sech^2(x).\label{eq7}
\en
representing the total volumetric displacement (which is conserved) of measure of $\eta_0(\xi)$.\footnote{It is worth noting that $\langle\eta_0\rangle=0$ for any wave motion that is
initiated from rest by defending the free surface from its
quiescent level (the state of static equilibrium); however, the
free-surface displacement in a laboratory wave tank, in which the
motion is initiated from a localized mound of water, is typically
measured from a depressed level surface that is {\it not} one of
static equilibrium.}

In (\ref{eq9b}) $\Ai(\chi)$ is an Airy function of the first kind, and is
characterized by a steeply rising wave front in $\xi
\gtrsim \tau^{\third}$ and by a slowly decaying, dispersive
wave train in $\xi \lesssim-\tau^{\third}$. Also, $\tau=\beta t/2$ is a slow varying time, where $\beta=(d/L)^2$ with $L$ representing a characteristic wavelength.

Figure (1) shows the three-dimensional plot of the real part of the solution (2.4), for $-15\leq\xi\leq10$, and $0.01\leq\tau\leq 5$.  As can be seen from this diagram the high frequency formation of solitary waves that occur for small values of $\tau$ tend to become more organized (lower frequency waves) as $\tau$ becomes larger.  Moreover, the solitary
waves totally decay, as we expect from the appearance of the Airy function in the solution (2.4). Note however, the `erratic' behavior of the solution for small values of $\tau$ is due to the fact that  the asymptotic solution of (2.4) is only valid for large 
$\tau$. 

\begin{figure}
   \begin{center}
\includegraphics[width=10cm]{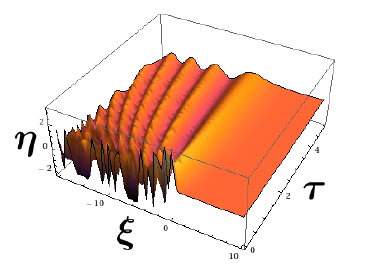}
   \end{center}
\caption{\footnotesize A prospective plot of the solution (2.4) for $-15\leq\xi\leq 10$
and  $0.01\leq\tau\leq 10$.}
\end{figure}

For values of $\xi\geq 0$, for all $\tau$, no solitons are detected.  Perhaps, formation of such solitons (or solitary waves in the present context) can be seen more readily in figure (2).  Figure (2) depicts the variation of $\eta$ with $\xi$ in the range $-15\leq\xi\leq 15$ for a fixed value of $\tau=0.01$ in  increments of $\Delta\xi=1.087$.

\begin{figure}
   \begin{center}
\includegraphics[width=14cm]{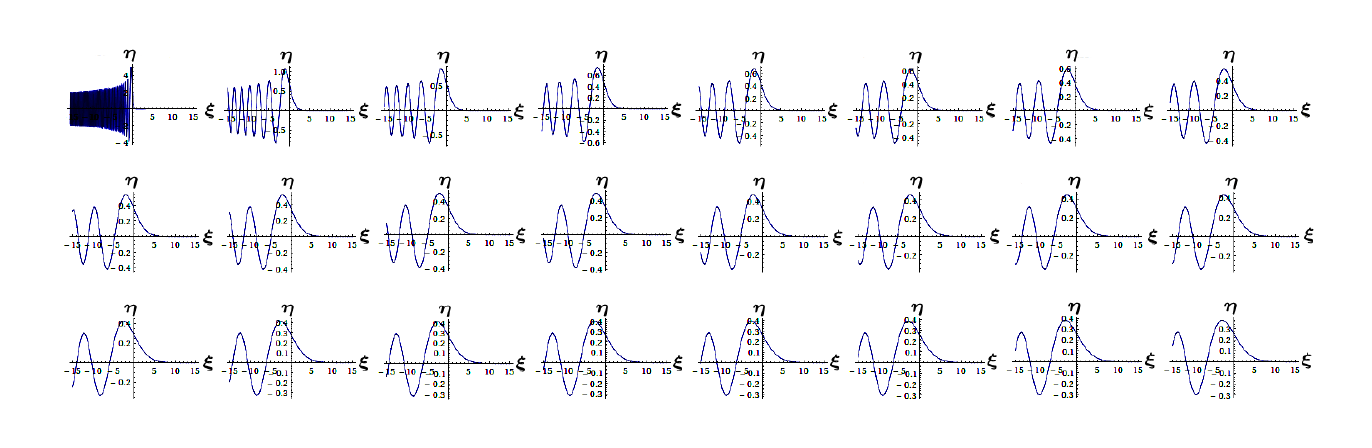}
   \end{center}
\caption{\footnotesize Spacial series of the solution (2.4) for $-15\leq\xi\leq 15$ at $\tau=0.01$.}
\end{figure}

In figures (3a) and (3b) we respectively show different prospective of the solution (2.4) for $\eta$ in the range $-25\leq\xi\leq10$ with $0.01\leq\tau\leq 5$, and for $-10\leq\xi\leq 10$ with $\tau$ in the same range as that of figure (3b). 

\begin{figure}
   \begin{center}
\includegraphics[width=6.5cm]{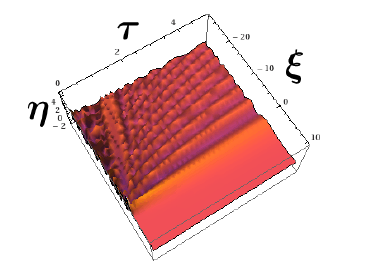}\quad\includegraphics[width=6.5cm]{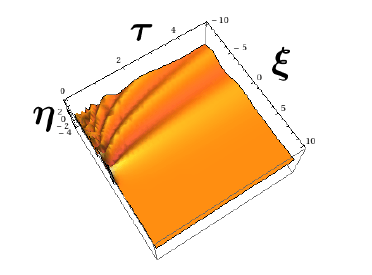}
   \end{center}
\caption{\footnotesize A prospective plot of the solution (2.4) for (a) $-25\leq\xi\leq 10$ and $0.01\leq\tau\leq 10$, left figure; (b) $-10\leq\xi\leq 10$
and  $0.01\leq\tau\leq 10$, right figure.}
\end{figure}

Following Miles [4], we remark that, it might
appear that the asymptotic solution of (\ref{eq4}) and (\ref{eq5}) for
$0<\eps\ll 1$ could be similarly characterized, however, it is
known that if $\langle\eta_0\rangle >0$ this solution comprises both a
decaying (as $\tau \uparrow\infty)$ wave train, which bears at least some
qualitative similarly to that predicted by linear theory, and a
soliton of the form
\be
\eta_1(\xi,\tau)=(\kappa^2/\eps)\,\mbox{sech}^2(\kappa\xi-\tf{4}{3}\kappa^3\tau+\delta),\label{eq10}
\en
which is fully evolved only for $\kappa\tau^{\third}\gg 1$. The
parameters $\kappa$ and $\delta$ can be determined by the
inverse-scattering method of Gardner {\it et al.} [7]; see also Whitham [8 \S 17.3], after letting $u=-2\eps\eta,
x=\xi$ and $t=\tau/3$ therein.

For the solution of the problem we follow Miles [4] and adopt the following path:

\noindent (i)  We solve the scattering problem posed by the one-dimensional
Schr\"{o}dinger equation
\be \{(d/dx)^2+k^2+2\eps\eta_0(x)\}\psi(x,k)=0\quad
(-\infty<x<\infty)\label{eq11} \en
subject to the radiation condition
\be 
\psi\sim e^{-ikx}+b(k)e^{ikx}\quad(x\uparrow\infty)\label{eq12} 
\en
and obtained the reflexion coefficient $b(k)$ over the continuous
spectrum $-\infty<k<\infty$.

\noindent (ii) Miles [4] stated that, the discrete eigenvalues, $k=i\kappa$, then are given by the poles of $b(k)$ on the {\it positive} imaginary axis of the
complex-$k$ plane, and he ruled out `false poles' by the restriction
that $\eta_0(x)$ the compact support.  But we argue that the restriction $\eps
\ll 1$ implies that there is at most one eigenvalue exists, whilst the condition $\langle\eta_0\rangle>0$
guarantees the existence of at least one such eigenvalue.  The
corresponding normalizing parameter is given by 
\be 
\gamma=-i{\rm Res}\{b(k),\,k=i\kappa\}.\label{eq13} 
\en
\noindent (iii)  Following Segur [10], we may show that the asymptotic
solution of the Gelfand-Levitan (or Marchenko) integral equation
is dominated by a soliton of the form (2.6) and that
\be 
\delta=\tf{1}{2}\ln(2\kappa/\gamma).\label{eq14} 
\en

Miles remarked that, the solution of this integral equation (see the next section) also yields a dispersive wave train, but an explicit representation thereof, even for
$\tau\rightarrow\infty$, is not available.

For comparison, we have also numerically integrated the KdV equation: (a), equation (2.1)
with $\eps=0.01$ subject to the initial condition that $\eta(\xi)=\sech^2(\xi)$ for $10\leq\xi\leq 10$ and
$0\leq\tau\leq 1.5$, and (b) the linearized KdV equation (2.2)
subject to the above conditions. For the numerical solution, in both cases, we have imposed the periodic boundary condition $\eta(-\infty,\tau)=\eta(\infty,\tau)$.

The results of the numerical integration for equations (2.1) and (2,2)
are shown in figures (4a) and (4b), respectively. From the result shown in figure (4a) we note that the salient features of the analytical solution of the linearized KdV equation
(\ref{eq4}) is captured fairly well. We remark that, in the presence of small nonlinearity, the multiple solitary waves that appeared in the  linearized equation (\ref{eq4}), is also present, and the two are very similar to each other.
However, we note for very small values of $\tau$ 
analytical and numerical solutions, for both linear and nonlinear KdV equations agree very well with each other. On the other hand, the results depicted in figure (4b) demonstrates the same behavior as that of the analytical solution (\ref{eq7}), provided $\xi\leq 0$. we emphasize, in the case when $\tau\gg 1$, the numerical solution of (2.1) and the analytical solution (2.4) are in good agreement with each other. 

\begin{figure}
   \begin{center}
\includegraphics[width=6.5cm]{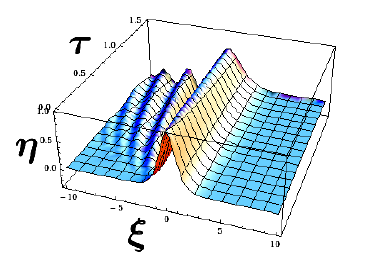}\quad\includegraphics[width=6.5cm]
{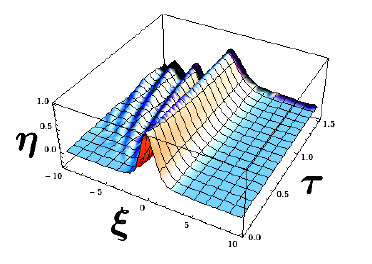}
   \end{center}
\caption{\footnotesize (a)  the numerical solution of equation (2.1), left figure; (b) the numerical solution of equation (2.2), right figure. Both for $-10\leq \xi\leq 10$ and $0\leq\tau\leq 1.5$.}
\end{figure}

\section{\sc Integral equation representation}

The scattering problem posed by (2.7) and (2.8) may be transformed
to an integral equation, using standard procedures used in the
quantum-scattering theory.  This integral equation may then be
solved by iteration, starting from (12) as a first approximation
to obtain
\be 
b(k)=-\eps\{ik+\eps\mathscr{N}(0)\}^{-1}\mathscr{N}(2k),\label{eq15} 
\en
where
\be
\mathscr{N}(k)=\int^{\infty}_{-\infty}e^{ikx}\eta_0(x)\,dx\label{eq16} 
\en
is the Fourier transform of $\eta_0(x)$, with errors of
$O(\eps^2)$ being implicit in both numerator and denominator of
(\ref{eq15}). Note that, there is a single discrete eigenvalue,
\be 
\kappa=\eps\langle\eta_0\rangle+O(\eps^2),\label{eq17} 
\en
if $\langle\eta_0\rangle>0$.  The corresponding approximation to the residue
of $-ib$ at $k=i\kappa$ is $\gamma=\kappa$, from which it follows
that
\be 
\delta=\tf{1}{2}\ln 2+O(\eps).\label{eq18} 
\en

In the next approximation $b(k)$ is given by
\be
b(k)=-\eps\left\{ik+\eps\int_{-\infty}^\infty e^{-ikx}f(x,-k)\eta_0(x)\,dx\right\}^{-1}
\int_{-\infty}^\infty e^{-ikx}f(x,k)\eta_0(x)\,dx\label{ieq3a}
\en 
where  $f(x,k)$ satisfies the integral equation
\be 
f(x,k)=e^{-ikx}+2\eps k^{-1}\int_x^\infty\sin[k(x-y)]f(y,k)\eta_0(y)\,dy\label{ieq3b}
\en 
Miles showed that (\ref{ieq3b}) may be approximated by
\be 
e^{ikx}f(x,k)=1+\eps(ik)^{-1}\int_x^\infty\left[e^{2ik(x-y)}-1\right]\eta_0(y)\,dy+O(\eps^2)\label{ieq3c}
\en 
Now, by substituting (\ref{ieq3c}) into (\ref{ieq3a}) we obtain the following result
\be 
b(k)=-\df{\eps\mathscr{N}(2k)+\eps^2(2ik)^{-1}\Int{-\infty}\infty\Int{-\infty}\infty
\left(e^{-2ikx}-e^{-2iky}\right)\eta_0(x)\eta_0(y)\sgn(x-y)\,dx\,dy}
{ik+\eps\mathscr{N}(0)-\eps^2(2ik)^{-1}\Int{-\infty}\infty\Int{-\infty}\infty\left(e^{2ik|x-y|}-1\right)\eta_0(x)\eta_0(y)\,dx\,dy}\no\\\label{ieq3d}
\en 
which is in  agreement with the equivalent result obtained  by Miles [4].
Thus, by evaluating the double integrals, for example
\be 
& &\Int{-\infty}\infty\Int{-\infty}\infty\left[e^{-2ik(x-y)}-1\right]\sech^2(x)\sech^2(y)\,dx\,dy\no\\ 
&=&2\Int{-\infty}\infty\sech^2(x)\left[\pi k e^{2ikx}\csch(\pi k)\right]\,dx\no\\
&=&4[\pi^2k^2\csch^2(\pi k)-1],\qquad\mbox{provided $-1<\Imag\{k\}<1$}\no 
\en 
etc., and transforming back, the analytical solution of the KdV can be obtained. Note that, the $O(\eps^2)$ analytical solution of the KdV equation may be expressed as
$$\eta(\xi,\tau)=2\eps\langle\phi_0^2\rangle\tau^{-\third}\Ai(\tau^{-\third}\xi)+
\langle\phi_0\rangle\tau^{-\third}\Ai'(\tau^{-\third}\xi)+O(\eps^3).$$
where $\phi_0(x)$ is given by (3.10) below.

It is to be noted that, the approximation (3.1), which implies the existence of a single
soliton if and only if $\langle\eta_0\rangle>0$, is uniformly valid for all
$k$ as $\eps\downarrow 0$ if $\langle\eta_0\rangle=O(1)$. A more significant example is
$\eta_0=\,\mbox{sech}^2(x)$, for which the exact solution of the
scattering problem yields
$$\kappa=\tf{1}{2}\{(1+8\eps)^{\half}-1\}$$ provided
$0<\eps\leq 1$ with
$\eps=\tf{1}{2}\mathscr{N}(\mathscr{N}+1)$ which corresponds to a pure soliton problem can be obtained
in the limiting approximation $\kappa=2\eps (\eps\downarrow 0)$ which would be in agreement
with (3.3).

It therefore follows that from (2.6) and (3.3) that the amplitude of the
soliton is $O(\eps a)$.  It also follows that the
normalized volumetric displacement of the soliton is
\be 
\langle\eta_1\rangle=2\langle\eta_0\rangle+O(\eps),\label{eq19} 
\en
and hence that the decaying wave-train component must have a
volumetric displacement $-\langle\eta_0\rangle$, rather than the value
$\langle\eta_0\rangle$ predicted by linear theory, in order to conserve the
initial value $\langle\eta_0\rangle$.
Note that, a single soliton also exists for when
$\langle\eta_0\rangle=0$, with
\be 
\kappa=2\eps^2\langle\phi_0^2\rangle+O(\eps^3),\quad
\phi_0(x)=\int^{\infty}_{\infty\,{\rm sgn}(x)}\eta_0(\zeta)\,d\zeta
\quad(\langle\eta_0\rangle=0),
\label{eq20} 
\en
where $\delta$ is given by (3.4).  The amplitude and volumetric
displacement of $\eta_1$ now are $\kappa^2/\eps=O(\eps^3)$
and $2\kappa/\eps=O(\eps)$, respectively, and therefore
are both negligible in the limit $\eps\downarrow 0$ with $\tau$ fixed;
nevertheless, $\eta_1$ dominates the solution in $\tau \gg
1/\kappa^3$.  

The solution for the next approximation may be reduced to
\be 
\eta(x)=2(1-i\eps k^{-1})e^{-ikx}\left[{}_2{\rm F}_1(1,ik;k+1;e^{-2x})-1\right]\label{eq21}
\en 
where ${}_2{\rm F}_1$ is the generalized hypergeometric function.  The plot of the real part $\eta(x)$ for $1\leq k \leq 4$ and $-10\leq x\leq 5$ is shown in figures (5a) and (5b) at two different prospective angles.  In this case we notice the erratic behavior 
seen in figure (1) is smoothed out and the solitary waves have smoother structure.  However, note the deviation and increase in amplitude of these waves in vicinity of $x=0$.  These waves can be interpreted as those which arrive at a shore, and as they approach the shallower water their amplitudes decrease and the waves dissipate.  We remark this feature is not consistent with those in a deeper water.

\begin{figure}
   \begin{center}
\includegraphics[width=6.5cm]{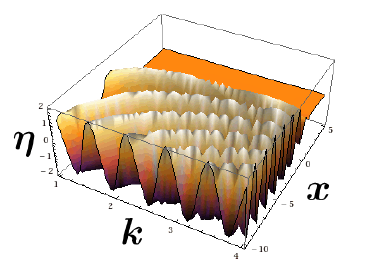}\quad\includegraphics[width=6.5cm]{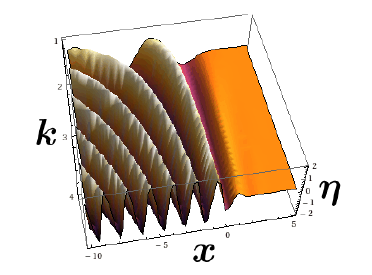}
   \end{center}
\caption{\footnotesize Two prospective plots of the solution (3.11) for $-10\leq x\leq 5$ and $1\leq k\leq 5$.}
\end{figure}

The complete account of the inverse scattering problem, which will also include the effect of water viscosity (see the next next section) and the analytical solution of the Marchenko integral equation will be reported in a subsequent paper.

At the conclusion of his paper, Miles makes the following crucial statement that `it perhaps should be emphasized that viscous
dissipation could prove more significant than nonlinearly over the
very long time intervals implied by $\eps^6\tau\gg 1.)$' With this in mind, we proceed to find an analytical solution to the viscous-KdV equation, which, to best of our knowledge, has not yet been discovered.  

\section{\sc Viscous KdV equation}

We consider propagation of the one-dimensional gravity wave whose amplitude is $a$, the wavelength is $L$ in a viscous liquid, with kinematic viscosity $\nu$, of uniform depth $d$.

We shall impose the following assumption that 
\be 
a\ll d\ll L\quad\mbox{and}\quad\delta\equiv (\nu L/c)^{\half},\label{vis1}
\en 
where $c=\sqrt{gd}$ is the wave speed and $\delta$ is the boundary-layer thickness. 

The equation governing the free-surface displacement $\eta$ and vertically averaged velocity $\langle u\rangle$, assuming that both $\eta$ and $\langle u\rangle$ are slowly varying functions in a reference frame moving with the speed $c$ satisfies the viscous-KdV equation
\be 
f_\tau+ff_\xi+\alpha f_{\xi\xi\xi}+\beta(\mathscr{D}f)_\xi=0\label{vis2}
\en 
where
\be 
f(\xi,\tau)=\eta(x,t)/a,\quad\xi=(x-ct)/L,\quad\tau=\df{3ac}{2dL}t\equiv\sigma t.\label{vis3}
\en 
The coefficients $\alpha$ and $\beta$ in (\ref{vis2}) are given by
\be 
\alpha=\df{d^3}{9aL^2}\quad\mbox{and}\quad\beta=\df{\delta}{3a},\label{vis4}
\en 
and
\be 
\mathscr{D}f=\int^{\infty}_{0}(\pi s)^{-\tf{1}{2}}f(\xi-s,\tau)\,ds.\label{vis5}
\en  

We note that if $\beta=0$ (\ref{vis2}) reverts back to the standard KdV equation.  Also note, equation (\ref{vis4}) provides for the case where only the bottom boundary layer exists.  However, to include side-wall boundary-layers of breath $b$ (say) we only need to modify $\beta$ by multiplying it by $1+2(d/b)$.

The diffusion operator $\mathscr{D}$ may be resolved, respectively, into dissipative and dispersive components $\mathscr{D}_1$ and $\mathscr{D}_2$ (say), according to 
\be 
\mathscr{D} f=\df{1}{2}\int^{\infty}_{-\infty}(\sgn s+1)|\pi s|^{-\tf{1}{2}} f(\xi-s,\tau)\,ds=(\mathscr{D}_1+\mathscr{D}_2) f\label{vis6}
\en 

We emphasize, $\mathscr{D}_1$ and $\mathscr{D}_2$ correspond to the odd and even components of the one-sided ($\equiv 0$ in $s<0$) function $(\pi s)^{-\tf{1}{2}}$.  We remark that $\mathscr{D}_1$ is the dissipative component of $\mathscr{D}$ which follows directly from spectral consideration. 

Let 
\be 
F(k,\tau)=\int^{\infty}_{-\infty} f(\xi,\tau)e^{-ik\xi}\, d\xi\label{vis7}
\en  
be the Fourier transform of $f$.  Then the transform of $\partial(\mathscr{D} f)/\partial\xi$ is $|\tf{1}{2}k|^{\tf{1}{2}}(1+i\sgn k)F$, in which the real and imaginary parts of it correspond to $\mathscr{D}_1$ and $\mathscr{D}_2$, respectively.  Thus, the Fourier transform of (\ref{vis2}) is given by
\be 
\left[\df{d}{dt}+\alpha(ik)^3+\beta|\tf{1}{2}k|^{\half}(1+i\sgn k)\right]F(k,\tau)+\df{ik}{4\pi}\int^{\infty}_{-\infty}F(\lambda,\tau)F(k-\lambda,\tau)\,d\lambda=0\,\,\,\label{vis8}
\en 
Multiplying (\ref{vis8}) through by $F(-k,\tau)$ and integrating over $-\infty<k<\infty$ yields the evolution equation 
\be 
\df{d}{dt}\int^{\infty}_{-\infty}|F|^2\,dk=-\beta\int^{\infty}_{-\infty}|2k|^{\tf{1}{2}}|F|^2\,dk,\label{vis9}
\en 
where $F^2$ is the power spectral density of $f$ and to which only the $\mathscr{D}_1$ component of $\mathscr{D}$ contributes.

From (\ref{vis9}) we obtain the first-order ordinary differential equation
\be 
\df{d|F|^2}{d\tau}+\beta|2k|^{\tf{1}{2}}|F|^2=\mbox{const.}\label{vis10}
\en 
We take the constant on the right-hand side of (\ref{vis10}) to be zero as there is only one initial condition remains to be satisfied.  For a solitary wave
$f(\xi,0)=\sech^2(\xi)$ whose Fourier transform is 
\be 
F(\xi,0)=\sqrt{\df{\pi}{2}} k \csch \left(\df{\pi k}{2}\right)\label{vis11}
\en 
The general solution of (\ref{vis10}) is given by
\be 
|F|^2=\mathscr{C}^2(k)\exp\{-\beta \tau|2k|^{\half}\}\label{vis12}
\en 

Applying the initial condition to evaluate $\mathscr{C}$ we obtain the solution of (\ref{vis12}) in the form
\be 
|F|=\sqrt{\df{\pi}{2}} k\csch\left(\df{\pi k}{2}\right)\exp\{-\tf{1}{2}\beta \tau|2k|^{\half}\}\label{vis13}
\en 
Hence, by taking the inverse Fourier transform of $F$ we obtain $f$ which is given by 
\be 
f=\df{1}{2\pi}\int^{\infty}_{-\infty}\sqrt{\df{\pi}{2}}
\df{k\exp\{-\tf{1}{2}\beta \tau|2k|^{\half}\}}{\sinh({\pi k/2})} e^{ik\xi}\,dk\label{vis14}
\en 

Next, using the expansion 
\be 
e^{-\sqrt{|\chi|}}=\sum^{\infty}_{n=0}\df{(-1)^n|\chi|^{\tf{n}{2}}}{n!}\label{vis15}
\en 
and substituting into (\ref{vis14}) and performing the integration, we obtain the closed formed solution
\be 
f(\xi,\tau)=\sech^2(\xi)+\sum^{\infty}_{n=1}\df{(-1)^n(n+2)\Gamma(\tf{n}{2})(\beta \tau)^n}{(2\pi)^{\tf{n}{2}+2}\Gamma(n)}\{\zeta(v_n,w)+\zeta(v_n,w^*)\}\label{vis16}
\en 
where $\zeta(v_n,w)$ is the generalized zeta function with $v_n=\df{n+4}{2}$ and $w=\df{1}{2}+i\df{\xi}{\pi}$, and the superscript $*$ indicates the complex conjugate.

Finally, restoring to the original variables the solution (\ref{vis16}) may be cast in form
\be 
\eta(x,t)=a\sech^2[(x-ct)/L]+a\sum^{\infty}_{n=1}\df{(-1)^n(n+2)\Gamma(\tf{n}{2})(\beta \sigma t)^n}{(2\pi)^{\tf{n}{2}+2}\Gamma(n)}\{\zeta(v_n,w)+\zeta(v_n,w^*)\}\no\\\label{vis16}
\en 
where now $w=\df{1}{2}+\df{i}{\pi L}(x-ct)$.

\begin{figure}
   \begin{center}
\includegraphics[width=11cm]{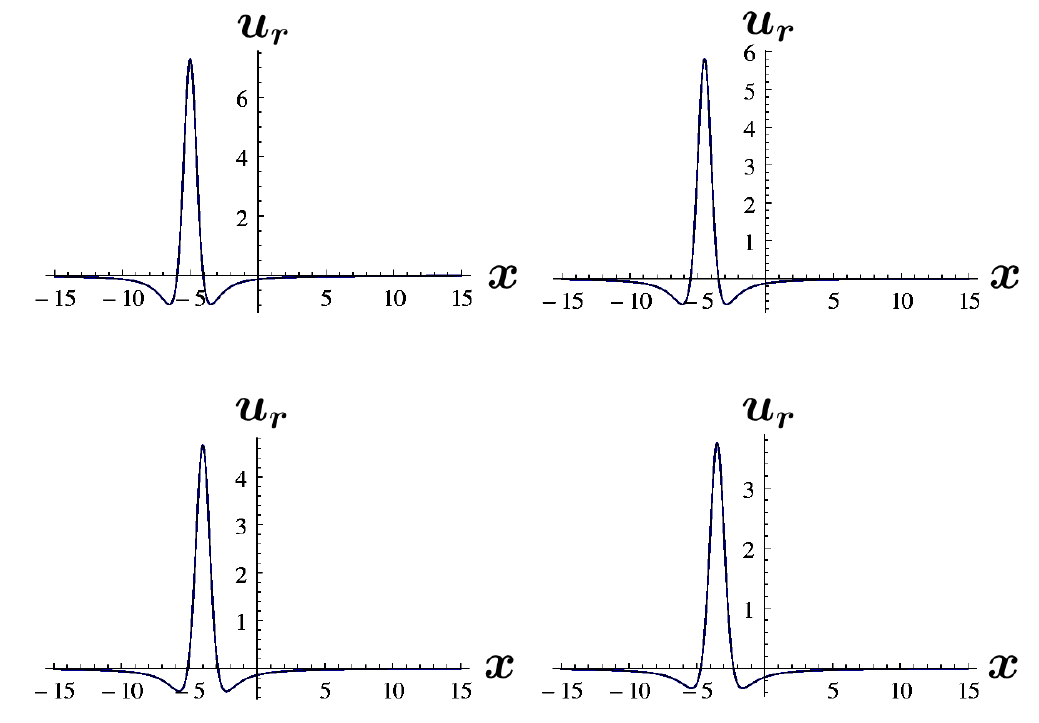}
\includegraphics[width=12cm]{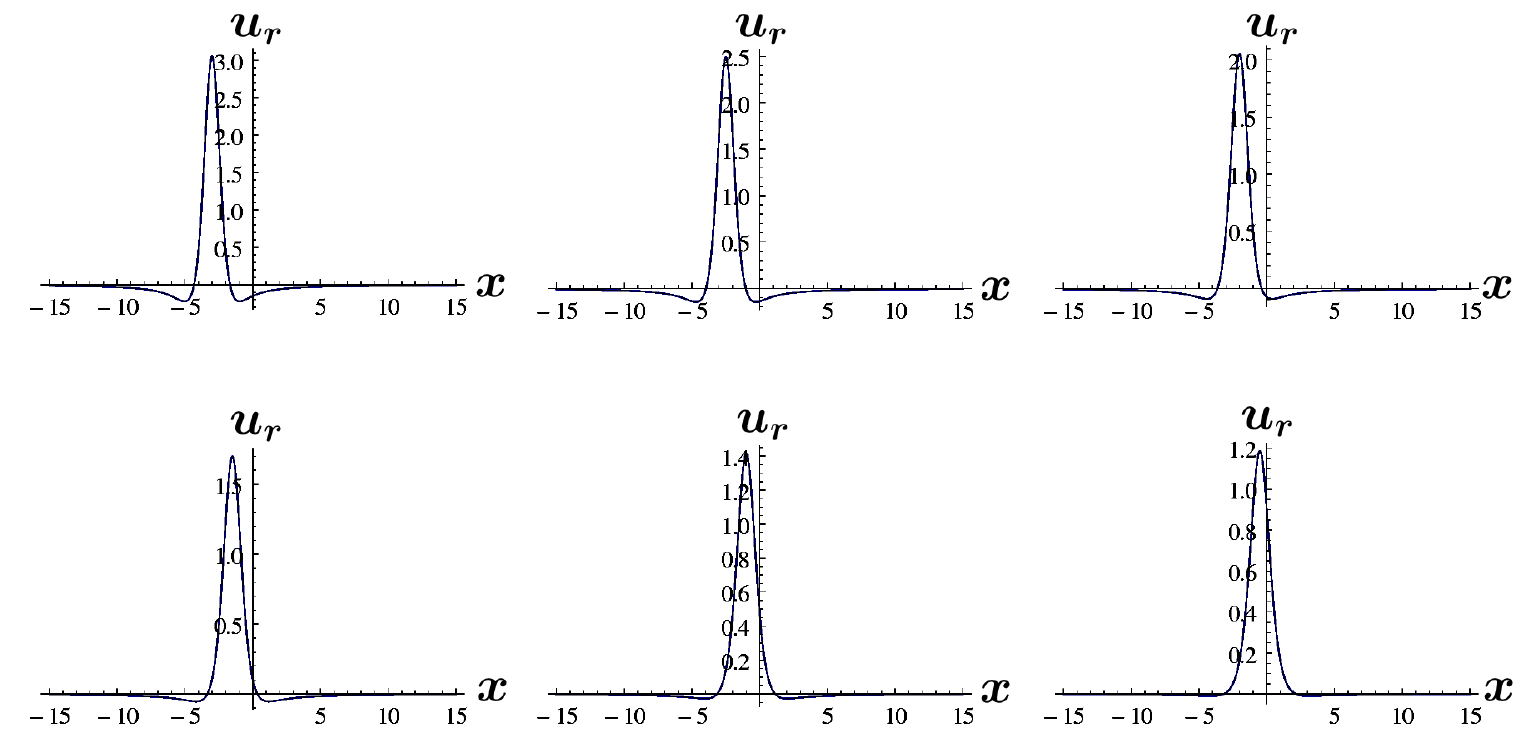}
\includegraphics[width=12cm]{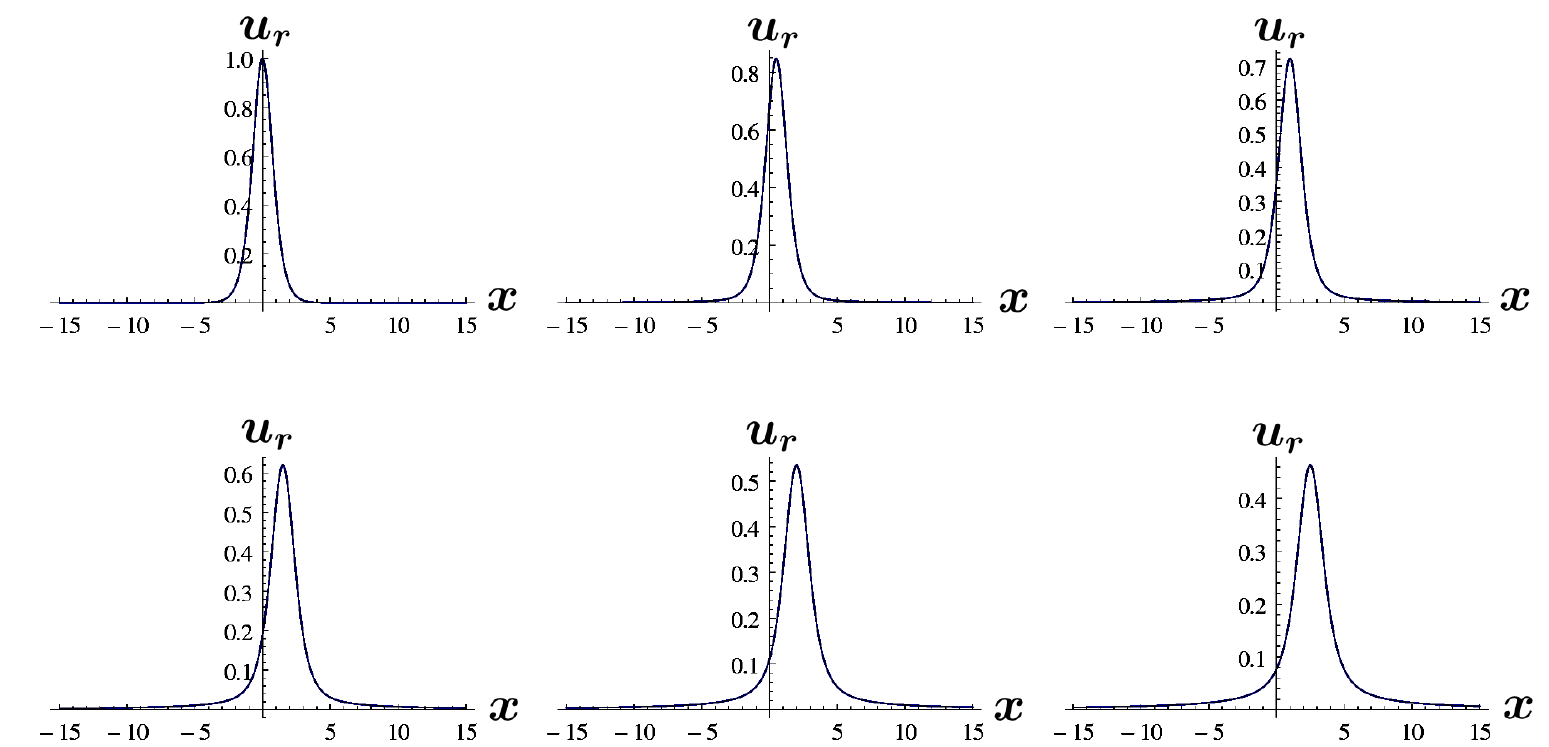}
   \end{center}
\caption{\footnotesize Time series solution of (4.17) for the range $-5\leq t\leq -1.061$.}
\end{figure}

\begin{figure}
   \begin{center}
\includegraphics[width=12cm]{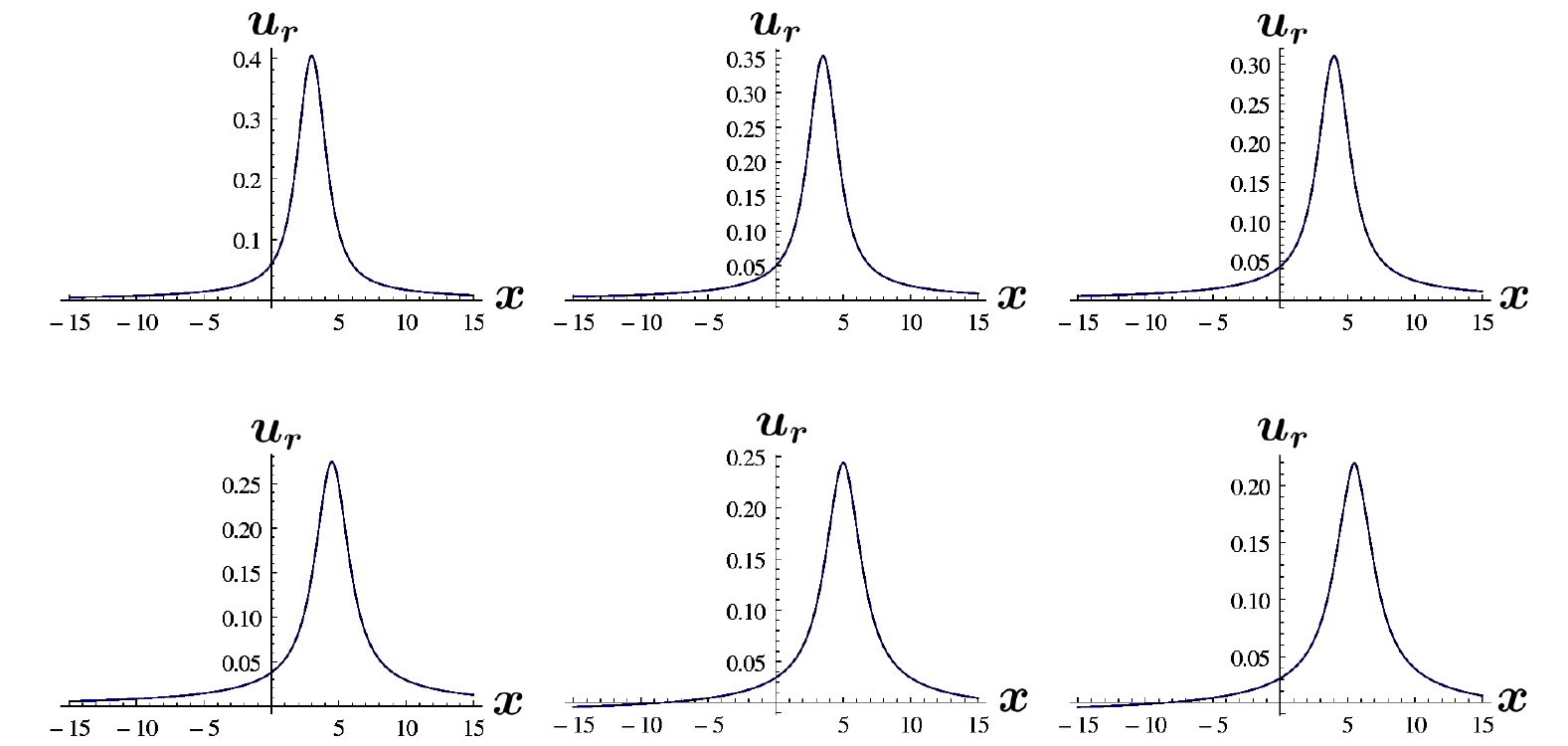}
\includegraphics[width=12cm]{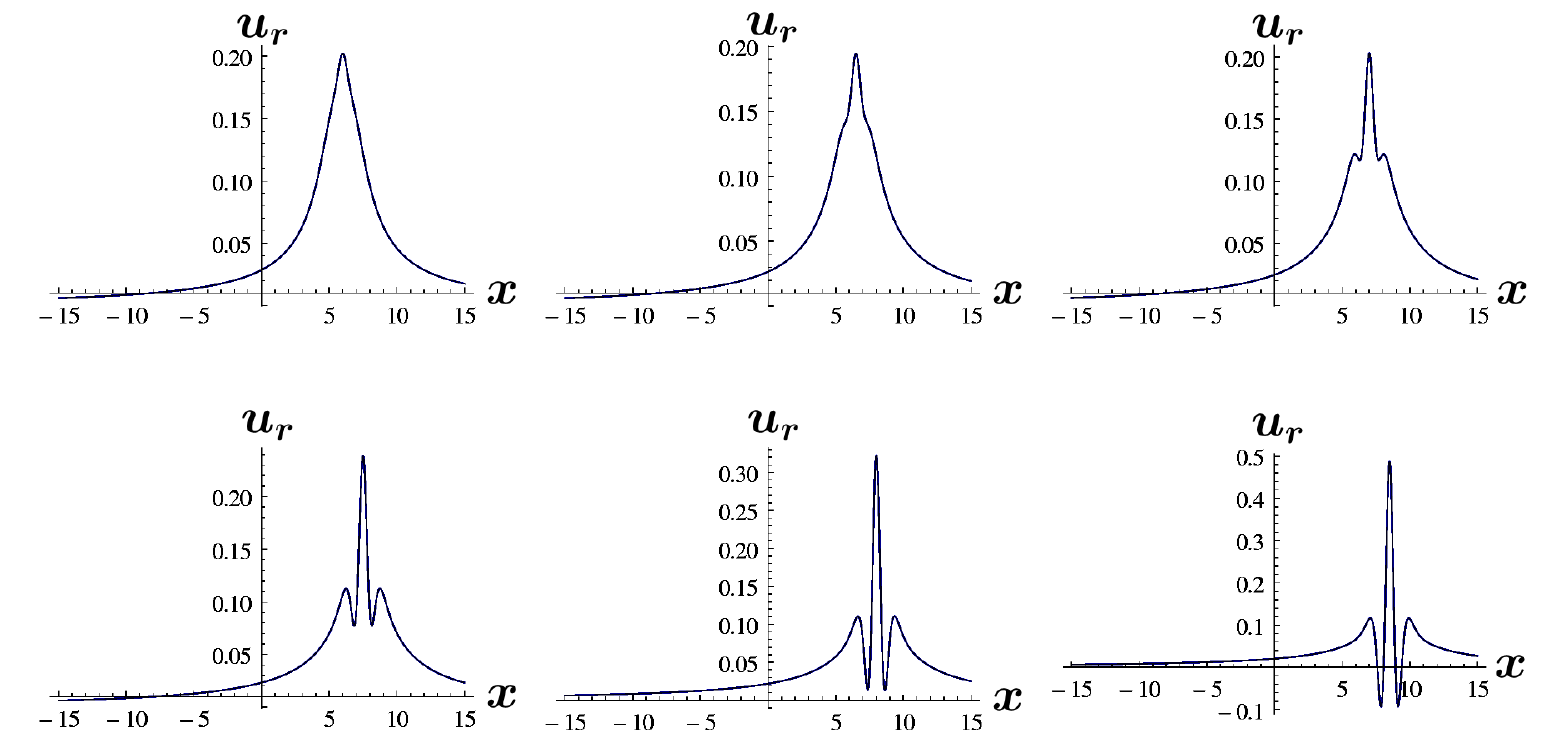}
\includegraphics[width=12cm]{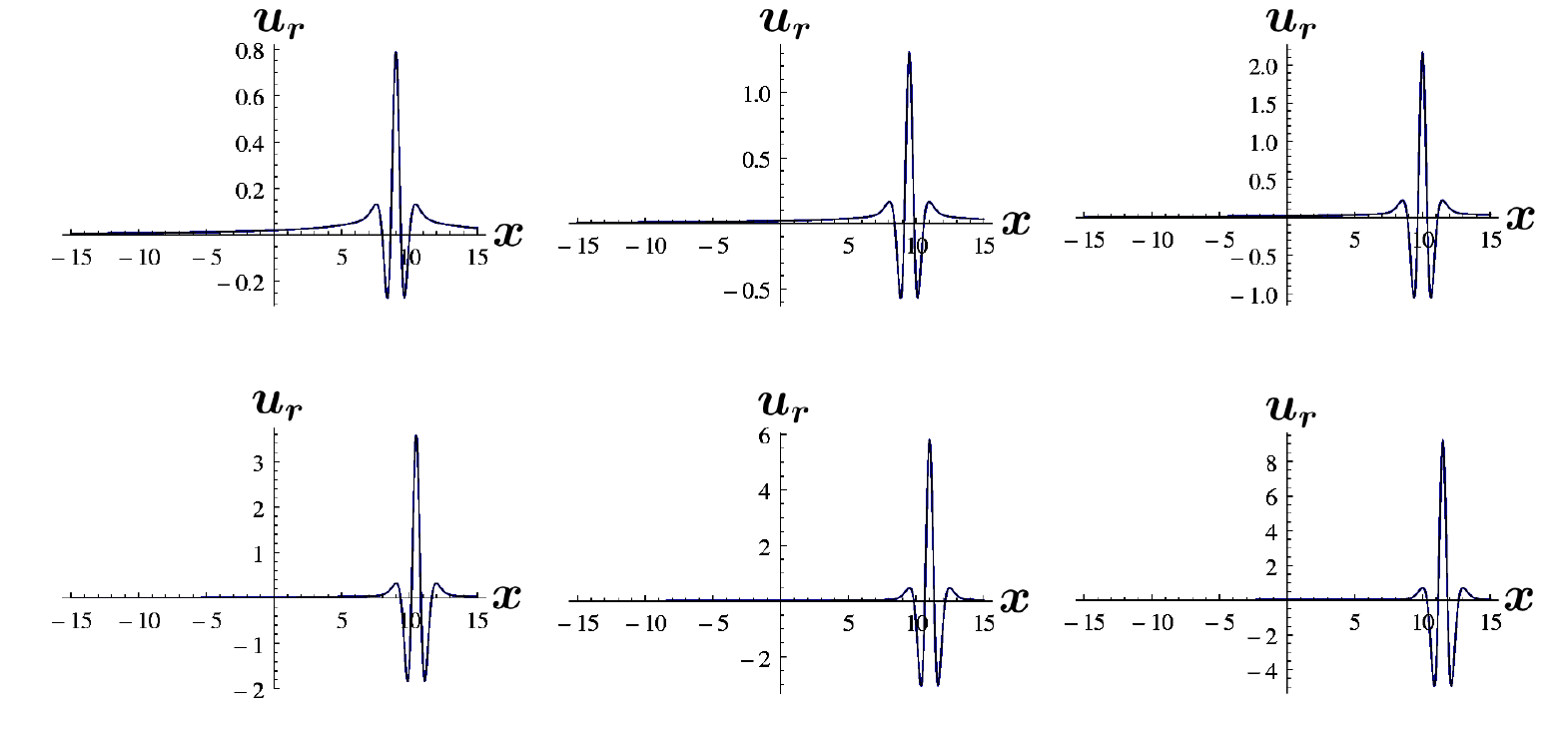}
   \end{center}
\caption{\footnotesize Time series solution of (4.17) for the range $-0.758\leq t\leq 5$.}

\end{figure}
\section{\sc Results and discussion}

For presentation of our results we shall denote the real part of the solution $f(\xi,\tau)$ by $u_r(x,t)\equiv\Real\{\eta(x,t)\}/a$, after restoring the transformed variables to the original ones.

In figure (6) and (7) we show the variation $u_r$ with $x$ in the range $-5\leq t\leq 5$ and time increments $\Delta t=0.303$.  From figure (6), we see that the solution of the viscous KdV equation demonstrates the formation of the Peregrine-type solution at $t=-5$.  As $t$ increases the amplitude of the soliton decreases and the eventually the solution at $t=0$, becomes $u_r=\sech^2[(x-ct)/L]$. We shall refer to this as the initial soliton.  This can be easily seen from the analytical solution (4.17).  Beyond the point $t=0$ the initial soliton moves to the right but its maximum amplitude decreases.  For  values of $t>0$ the initial soliton widens up, although its amplitude decreases to about $\tf{1}{5}$ of the amplitude of the initial soliton.  As $t$ increases further, a secondary soliton begins to be formed around the peak of the  primary soliton. This secondary soliton becomes more pronounced for even larger values of $t$.  At $t=2.272$ the secondary soliton that is formed on both sides of the initial soliton begins to collapse, but its maximum amplitude of the primary soliton increases and it also travels farther to the right. 

The behavior of the soliton is now consistent with a breather.  Moreover, the soliton once again becomes like that of Peregrine soliton mentioned above.   Note that, at large values of $t$ the solitons become narrower and steeper.  However, we must emphasize the validity of solution (4.17) lies within the convergence of the infinite series (4.15).  In the present problem it appears that the solution converges in the range $-5\leq t \leq 5$  (with respect to the physical variables).  We must stress that proving the absolute convergence of the solution (4.17) is relatively hard.  This is because the solution contains the generalized zeta function of complex argument.

As can be seen from figure (8), which shows the solution (4.17) in the range $-15\leq x\leq 15$ for $0\leq t \leq 5$, the primary soliton propagates with decaying amplitude and eventually bifurcates as $t\rightarrow 5$.  We also note the propagation of the soliton, though decaying in amplitude, shifts more to the positive $x$-direction.  It can also be observed that this decaying soliton is not completely smooth.  

Figure (9a) shows the solution of (4.17) in the range $-10\leq x\leq 10$ but for $-5\leq t\leq 5$.  As can be seen from this figure the high-amplitude soliton decays.  At $t=0$ the solution is simply $\sech^2(x)$, but for large values of $t\sim O(4)$ the bifurcation in the solitons very visible.  However, for the same $x$-range as that of figure (9a), we see from figure (9b) the bifurcated soliton begins to grow in amplitude rapidly around $t\sim O(9)$.  This may be attributed to the fact that in the vicinity of this value of $t$ the solution (4.17) begins to diverge.  

The remark we just made warrants further in-depth investigation regarding the convergence of the solution (4.17), in particular the validity of its rate of convergence with respect to time, $t$. 

One alternative study can be conducted by transforming the viscous-KdV equation to a viscous-nonlinear Schr\"odinger (VNLS) equation of the form [9]\footnote{Recently Sajjadi has found  two possible solution to the equation (\ref{sch1}) which will be reported shortly.}
\be 
A_{\tau}+i\beta A_{\xi\xi}+i\gamma|A|^2A+\alpha(\mathscr{D}A)_{\xi}=0\label{sch1}
\en 
using $$\eta(x,t)=\eps A(\xi,\tau)e^{i\theta_0}+\eps^2\left[B(\xi,\tau)e^{2i\theta_0}+C(\xi,\tau)\right]+{\rm c.c.}$$ where 
$$\theta_0=k_0x-\omega_0t,\quad{\rm with}\quad \xi=\eps(x-ct),\quad{\rm and}\quad\tau=\eps^2 t$$. 
 
However, this formulation only includes the dissipative term (but not dispersive).  Furthermore, as was shown by Sajjadi {\it et al.} [10], there are many solutions to the NLS equation, depending on initial and boundary conditions, but not every solution are representative of a physical case, or relevant to the problem(s) under consideration.  

We, thus expect that with inclusion of the viscous term, there would exist some 
non-physical solutions which must be relegated.  Following the work of Karjanto and Tiong [9] we obtain the same nonlinear dispersion relation (provided the dispersive term is excluded)
$$\omega=-\beta k^2+\gamma|A|^2+\alpha k(1-i)\{\eps/[2(k_0+\eps k)]\}^{\Half}$$
This result (which the details are omitted here) is in agreement with Karjanto and Tiong [9] equation (2.9).  We remark that the above VNSL equation (\ref{sch1}) is very useful for studying modulational instability in the presence of viscosity [9]. Moreover, the theory presented by Karjanto and Tiong [9] may be extended to include the dispersive term in addition to the dissipative term.

Finally, we remark that the result of such a study may shed more light in explaining the damping of the Benjamin-Feir instability by viscosity, such as that investigated by Segur
{\it et al.} [11], especially in problems related to wind generated waves which Sajjadi {\it et al.} [10] has recently reported but with the inclusion of viscosity in the water.

\begin{figure}
   \begin{center}
\includegraphics[width=6cm]{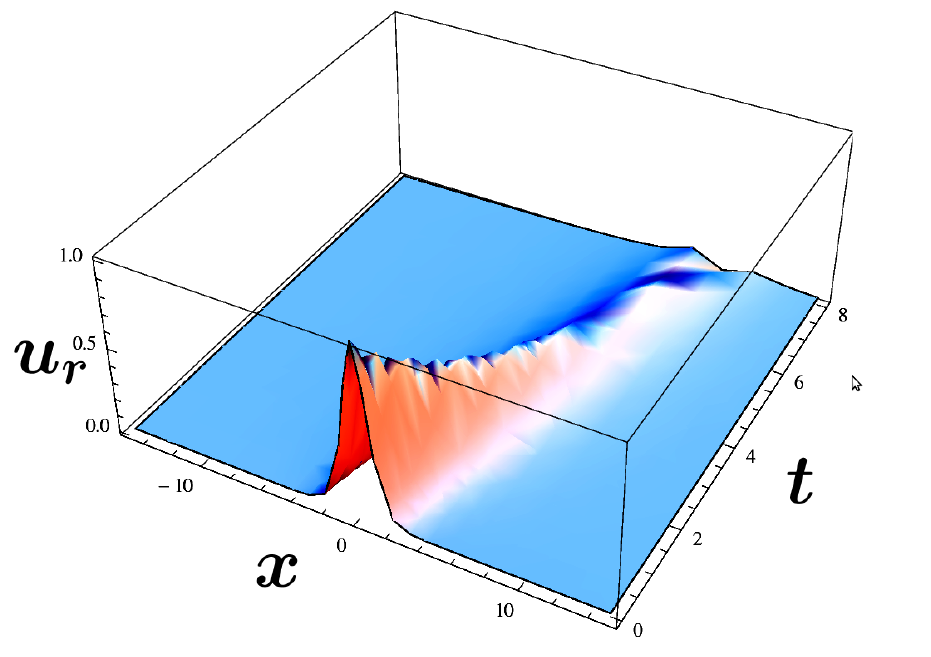}\quad\includegraphics[width=6cm]{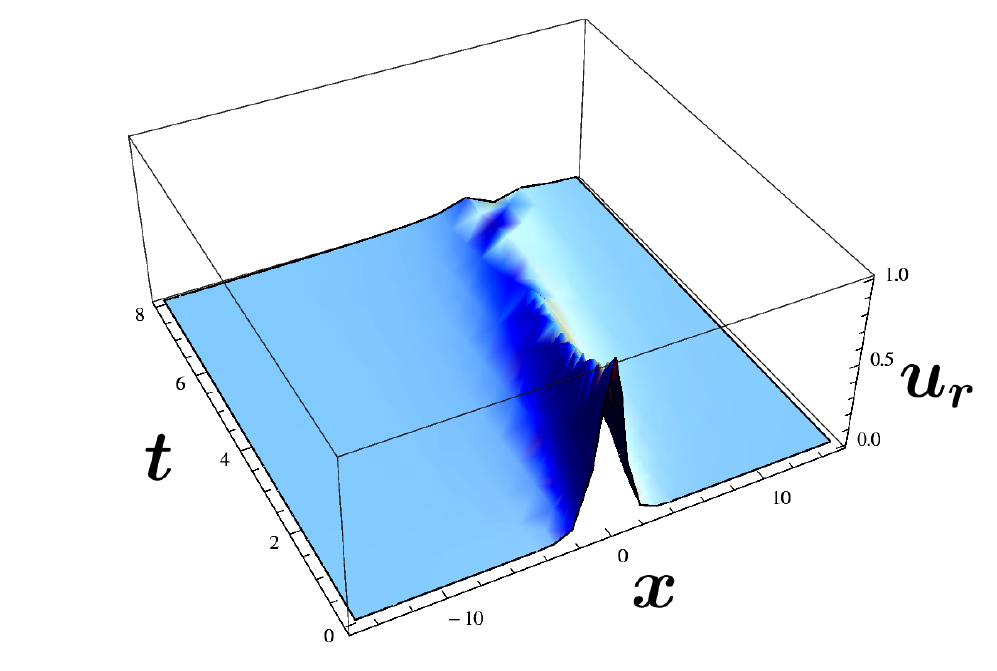}
   \end{center}
\caption{\footnotesize Two prospective plots of the solution (4.17) as a function of $x$ and $t$ for $-15\leq x\leq 15$ and $0\leq t\leq 8$.}
\end{figure}

\begin{figure}
   \begin{center}
\includegraphics[width=6.5cm]{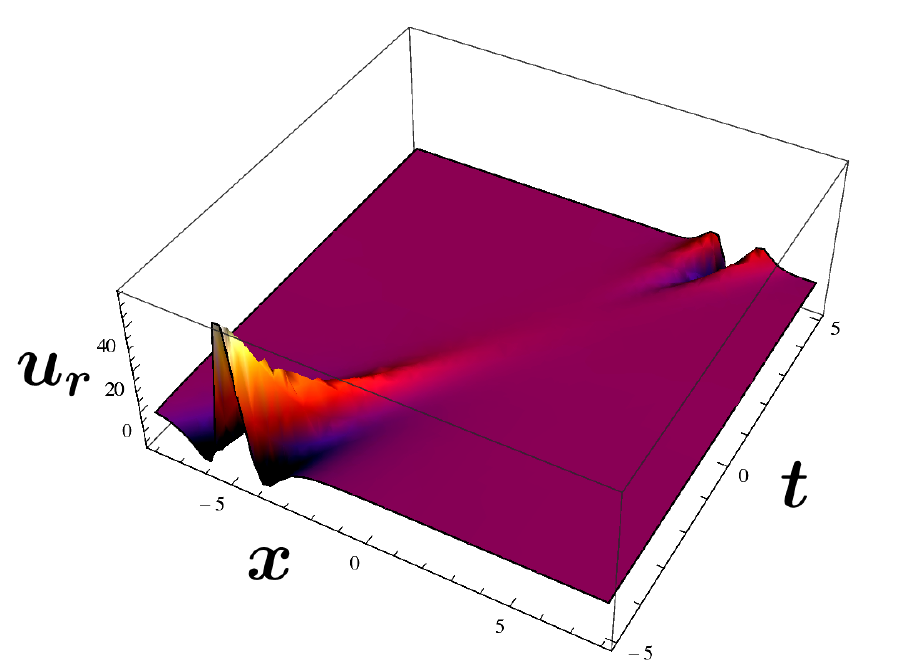}\quad\includegraphics[width=6.5cm]{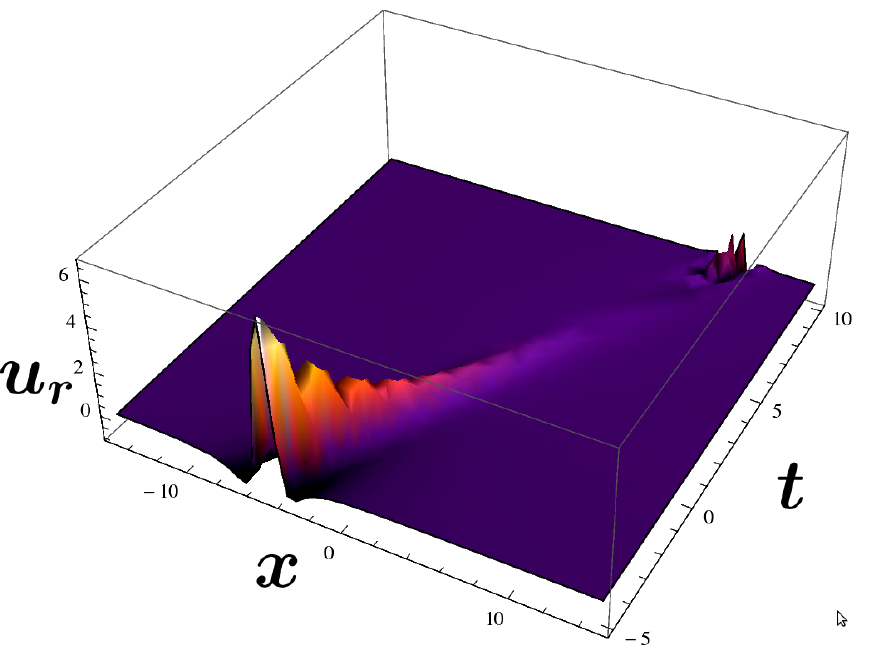}
   \end{center}
\caption{\footnotesize Two prospective plots of the solution (4.17) as a function of $x$ and $t$, (a) for $-8\leq x\leq 8$ and $-5\leq t\leq 5$, left figure; (b) for $-13\leq x\leq 13$ and $-5\leq t\leq 10$, right figure.}
\end{figure}

\section{\sc Conclusions}
In this paper we revisited (a) Miles [4] theory of the evolution of a solitary wave with very weak nonlinearity; and (b) the viscous counterpart of (a), except the nonlinearity is no longer weak.  In both cases, we have considered a one-dimensional gravity wave of amplitude $a$ and wavelength $L$ in (a) inviscid and (b) viscous water of uniform depth $d$.  In the case (a) the problem reduces to the regular KdV equation, which we solved asymptotically in the case (i) when the nonlinearity is very weak and (ii) where nonlinearity is totally omitted.  For the case (ii) the solution is expressed in terms of Airy function, but with the restriction that the solution is only valid for `slow time' $\tau=\tf{1}{2}(d/L)^2(gd)^{\tf{1}{2}}\uparrow 0$.  For comparison, we have also solved the KdV equation for both cases (i) and (ii) numerically.  In both cases, the measure of nonlinearity is given by $\eps=3aL^2/4d^3\ll 1$.  In contrast for the case (b) in which the KdV is modified by viscosity, the analytical solution is expressed in terms of generalized zeta function.  The intriguing results obtained for the  case (b) show the formation of Peregrine soliton which propagates and eventually tends to the regular $\sech^2(x)$ soliton in the vicinity of $t=0$.  For $t>0$ the regular soliton widens and eventually two secondary solitons are formed around the peak of the primary one.  Then for $t\gg 1$ the solitons collapse and become narrower, and finally bifurcates to two individual solitons. 

\section*{\sc Note added at the proof}
Recently we have observed photographs which supports the conclusion that the solitons (or in the present case the solitary waves) bifurcates to two individual solitons from the aerial photographs taken from a helicopter by Japanese film crew flying over the Japan tsunami of 2011. These photographs are depicted in figures (10).

\section*{\sc Acknowledgement}

We would like to thank Professor Thanasis Fokas, of Department of Applied Mathematics and Theoretical Physics, University of Cambridge, for reading the manuscript and make valuable comments. 

\begin{figure}
   \begin{center}
\includegraphics[width=6.5cm]{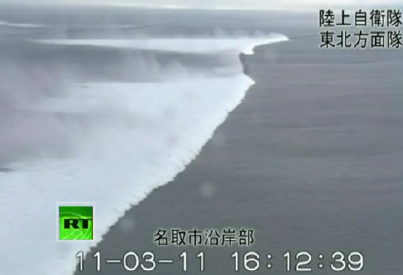}\quad\includegraphics[width=6.5cm]{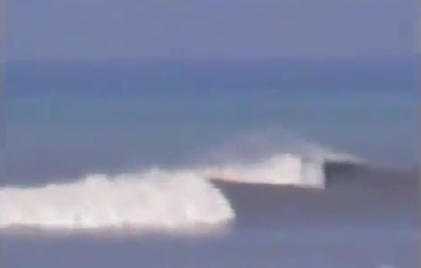}
   \end{center}
\caption{\footnotesize Two aerial photographs taken from a helicopter by Japanese film crew flying over the Japan tsunami of 2011.}
\end{figure} 

\section*{\sc References}

\parindent=0pt
\parskip=\smallskipamount
\baselineskip=\normalbaselineskip \hangindent=2em
\def\hang{\hangindent=2em \hangafter=1}
\def\\{\hfil\break}

\hang [1] W. Chester, Resonant oscillations of water waves. Proceedings of the Royal Society A, 306, 5, 1968.

\hang [2] E. Ott and R.N. Sudan, Damping of solitary waves. I. Theory. Physics of Fluids, 13, 1432, 1970.

\hang [3] J.W. Miles, Korteweg-de Vries equation modified by viscosity. Physics of Fluids, 19, 1063, 1976.

\hang [4] J.W. Miles,  The Ursell paradox. In proceedings of Geofluid dynamical wave mathematics, University of Washington, pp. 103-110, 1977.

\hang [5] C.G. Koop and G. Butler, An investigation of internal solitary waves in a two-fluid system. Journal of Fluid Mechanics, 112, 225, 1981.

\hang [6] K.P. Das, A Korteweg-de Vries equation modified by viscosity for waves in a channel of uniform but arbitrary cross section. Physics of Fluids, 28, 770, 1985.

\hang [7] C.S. Gardner, J.M. Greene, M.D. Kruskal, and M.M. Robert. Method for solving the Korteweg-deVries equation. Physical Review Letters, 19, 1095, 1967.

\hang [8] G.B. Whitham, Linear and nonlinear waves (Vol. 42). John Wiley and Sons, 2011.

\hang [9] N. Karjanto and K.M. Tiong, Stability of the NLS equation with viscosity effect. Journal of Applied Mathematics, 2011.

\hang [10] S.G. Sajjadi, S.C. Mancas and F. Drullion, Formation of three-dimensional surface waves on deep-water using elliptic solutions of nonlinear Schr\"odinger equation.
Advances and Applications in Fluid Mechanics, 18, 81, 2015.

\hang [11] H. Segur, D. Henderson, J. Carter, J. Hammack, C.M. Li, D. Pheiff and K. Socha, Stabilizing the Benjamin-Feir instability. Journal of Fluid Mechanics, 539, 229, 2005.

\end{document}